\documentstyle[11pt,aaspp4,epsfig]{article}  
\def\gray{$\gamma$-ray\ }
\def\grays{$\gamma$-rays\ }
\def\BC{B/C}
\def\Berat{$^{10}$Be/$\,^9$Be}
\def\Dpp{D_{pp}}
\def\Dxx{D_{xx}}
\def\ddp{{\partial\over\partial p}}

\def\fwb{65mm}

\lefthead{STRONG \& MOSKALENKO}
\righthead{GALACTIC COSMIC RAYS AND GAMMA RAYS: A UNIFIED APPROACH
}

\begin{document}

\renewcommand{\thefootnote}{\fnsymbol{footnote}}

\title{
Galactic cosmic rays and gamma rays: a unified approach%
\footnote[1]{\it
   Contribution to a book ``Topics in Cosmic Ray Astrophysics'',
   ed.~M.A.DuVernois (NY: Nova Scientific), 1999 }
}

\renewcommand{\thefootnote}{\arabic{footnote}}

\author{Andrew W.~Strong\altaffilmark{1} and 
   Igor V.~Moskalenko\altaffilmark{1,2}}

\affil{\altaffilmark{1}Max-Planck-Institut f\"ur extraterrestrische Physik,
   Postfach 1603, D-85740 Garching, Germany}
\affil{\altaffilmark{2}Institute for Nuclear Physics, 
   M.V.Lomonosov Moscow State University, 119 899 Moscow, Russia}

\authoremail{aws@mpe.mpg.de; imos@mpe.mpg.de}

\begin{abstract}
We are constructing a model which aims to reproduce
observational data of many kinds related to cosmic-ray origin and
propagation: direct measurements of nuclei, antiprotons, electrons and
positrons, $\gamma$-rays, and synchrotron radiation. These data provide
many independent constraints on any model.
Propagation of primary and secondary nucleons, primary
and secondary electrons and positrons are calculated self-consistently.
Fragmentation and energy losses are computed using realistic
distributions for the interstellar gas and radiation fields, and
diffusive reacceleration is also incorporated.  The models are adjusted
to agree with the observed cosmic-ray \BC\ and \Berat\ ratios.

Our main results include evaluation of diffusion/convection and
reacceleration models, estimates of the halo size, calculations of the
interstellar positron and antiproton spectra, evaluation of alternative
hypotheses of hard nucleon and hard electron interstellar spectra, and
computation of the Galactic diffuse \gray emission.

\end{abstract}

\section{Introduction} \label{introduction}

We have developed a model which aims to reproduce self-consistently
observational data of many kinds related to cosmic-ray origin and
propagation: direct measurements of nuclei, antiprotons, electrons and
positrons, $\gamma$-rays, and synchrotron radiation. These data provide
many independent constraints on any model and our approach is able to
take advantage of this since it must be consistent with all types of
observation.

A numerical method and corresponding computer code (GALPROP) for the
calculation of Galactic cosmic-ray propagation in 3D has been
developed.  The basic spatial propagation mechanisms are
(momentum-dependent) diffusion and convection, while in momentum space
energy loss and diffusive reacceleration are treated.  Fragmentation
and energy losses are computed using realistic distributions for the
interstellar gas and radiation fields.  The code is sufficiently
flexible that it can be extended to include new aspects as required.
The basic procedure is first to obtain a set of propagation parameters
which reproduce the cosmic ray \BC\ and \Berat\ ratios; the same
propagation conditions are then applied to primary electrons. Gamma-ray
and synchrotron emission are then evaluated with the same model.

Our approach is not intended to perform detailed source abundance
calculations with a large network of reactions, which is still best
done with the path-length distribution approach (see e.g.
\cite{DuVernois96} and references therein). Instead we use just the
principal progenitors and weighted cross sections based on the observed
cosmic-ray abundances (see \cite{Webber92}).  The \BC\ data is used since
it is the most accurately measured ratio covering a wide energy range
and having well established cross sections.  A re-evaluation of the
halo size is desirable since new \Berat\ data are available from
Ulysses with better statistics than previously.

Preliminary results were presented in Strong \& Moskalenko (1997)
(hereafter \cite{SM87}) and full results for protons,
Helium, positrons, and electrons in Moskalenko \& Strong (1998a)
(hereafter \cite{MS98a}).  Evaluation of the \BC\ and
\Berat\ ratios, evaluation of diffusion/convection and reacceleration
models, and setting of limits on the halo size, as well as full details
of the numerical method and energy losses for nucleons and electrons
are summarized in Strong \& Moskalenko (1998) (hereafter \cite{SM98}).
Evaluation of  antiprotrons in connection with diffuse Galactic
$\gamma$-rays and interstellar nucleon spectrum are given in
Moskalenko, Strong, \& Reimer (1998) (hereafter \cite{MSR98}).  For a
recent discussion of diffuse Galactic continuum \grays and synchrotron
emission in the context of this approach see Strong, Moskalenko, \&
Reimer (1998) (hereafter \cite{SMR98}) and Moskalenko \& Strong
(1998d).

For interested users our model is available in the public domain on the
World Wide Web ({\it
http://www.gamma.mpe-garching.mpg.de/$\sim$aws/aws.html})

\section{Motivation}

It was pointed out many years ago (see \cite{Ginzburg80},
\cite{Berezinskii90}) that the interpretation of radioactive cosmic-ray
nuclei is model-dependent and in particular that halo models lead to a
quite different physical picture from homogeneous models.  The latter
show simply a rather lower average matter density than the local
Galactic hydrogen (e.g., \cite{SimpsonGarcia88,Lukasiak94a}), but do
not lead to a meaningful estimate of the size of the confinement
region, and the correponding cosmic-ray `lifetime' is model-dependent.
In such treatments the lifetime is combined with the grammage to yield
an `average density'.  For example Lukasiak et al. (1994a) find an
`average density' of 0.28 cm$^{-3}$ compared to the local interstellar
value of about 1 cm$^{-3}$, indicating a $z$-extent of less than 1 kpc
compared to the several kpc found in diffusive halo models. 
Our model includes spatial dimensions as a basic
element, and so these issues are automatically addressed.

The possible r\^ole of convection was shown by Jokipii (1976), and
Jones (1979) pointed out its effect on the energy-dependence of the
secondary/primary ratio.  Recent papers give estimates for the halo
size and limits on convection based on existing calculations (e.g.,
\cite{Webber92}, \cite{WebberSoutoul98}), and we attempt to improve on
these models with a more detailed treatment.

Previous approaches to the spatial nucleon propagation problem have
been mainly analytical: Jones (1979), Freedman et al. (1980),
Berezinskii et al. (1990), Webber, Lee, \& Gupta (1992), Bloemen et al.
(1993), and Ptuskin \& Soutoul (1998) treated diffusion/convection
models in this way.  Bloemen et al. (1993) used the `grammage'
formulation rather than the explicit isotope ratios, and their
propagation equation implicitly assumes identical distributions of
primary and secondary source functions.  These papers did not attempt
to fit the low-energy ($<1$ GeV/nucleon) \BC\ data (which we will show
leads to problems) and also did not consider reacceleration.  It is
clear than an analytical treatment quickly becomes limited as soon as
more realistic models are desired, and this is the main justification
for the numerical approach. The case of electrons and positrons is even
more intractable analytically, although fairly general cases have been
treated (\cite{Lerche82}).  Recently Porter \& Protheroe (1997) made
use of a Monte-Carlo method for electrons, with propagation in the
$z$-direction only.  This method would be very time-consuming for 2- or
3-D cases. Our method, using numerical solution of the propagation
equation, is a practical alternative.

Reacceleration has previously been handled using leaky-box calculations
(\cite{Letaw93,SeoPtuskin94,HeinbachSimon95}); this has the advantage
of allowing a full reaction network to be used (far beyond what is
possible in the present approach), but suffers from the usual
limitations of leaky-box models, especially concerning radioactive
nuclei, which were not included in these treatments.  Our simplified
reaction network is necessary because of the added spatial dimensions,
but we believe it is fully sufficient for our purpose, since we are not
attempting to derive a comprehensive isotopic composition.  A more
complex reaction scheme would not change our conclusions.

\section{Description of the models} \label{Description}

The models are three dimensional with cylindrical symmetry in the
Galaxy, and the basic coordinates are $(R,z,p)$, where $R$ is
Galactocentric radius, $z$ is the distance from the Galactic plane, and
$p$ is the total particle momentum.  The distance from the Sun to the
Galactic centre is taken as $R_\odot=8.5$ kpc.  In the models the
propagation region is bounded by $R=R_h$, $z=\pm z_h$ beyond which free
escape is assumed. We take $R_h=30$ kpc. The range $z_h=1-20$ kpc is
considered.  For a given $z_h$ the diffusion coefficient as a function
of momentum is determined by \BC\ for the case of no reacceleration; if
reacceleration is assumed then the reacceleration strength (related to
the Alfv\'en speed) is constrained by the energy-dependence of \BC.
The spatial diffusion coefficient for the case of no reacceleration is
taken as $\Dxx = \beta D_0(\rho/\rho_0)^{\delta_1}$ below rigidity
$\rho_0$, $\beta D_0(\rho/\rho_0)^{\delta_2}$ above rigidity $\rho_0$,
where the factor $\beta$ ($= v/c$) is a natural consequence of a
random-walk process.  Since the introduction of a sharp break in $\Dxx$
is a contrived procedure which is adopted just to fit
\BC\ at all energies, we also consider the case $\delta_1=\delta_2$,
i.e. no break, in order to investigate the possibility of reproducing
the data in a physically simpler way.  The convection velocity (in
$z$-direction only) $V(z)$ is assumed to increase linearly with
distance from the plane ($V>0$ for $z>0$, $V<0$ for $z<0$, and
$dV/dz>0$ for all $z$); this implies a constant adiabatic energy loss.
The linear form for $V(z)$ is suggested by cosmic-ray driven MHD wind
models (e.g., \cite{Zirakashvili96}).

We include diffusive reacceleration since some stochastic
reacceleration is inevitable, and it provides a natural mechanism to
reproduce the energy dependence of the \BC\ ratio without an {\it ad
hoc} form for the diffusion coefficient
(\cite{Letaw93,SeoPtuskin94,HeinbachSimon95,SimonHeinbach96}).  The
spatial diffusion coefficient for the case of reacceleration assumes a
Kolmogorov spectrum of weak MHD turbulence so $\Dxx=\beta
D_0(\rho/\rho_0)^\delta$ with $\delta=1/3$ for all rigidities.
For this case the momentum-space diffusion coefficient $D_{pp}$ is
related to the spatial coefficient using the formula given by Seo \&
Ptuskin (1994), and Berezinskii et al.\ (1990)
\begin{equation}
\label{2.1}
\Dpp\Dxx = {4 p^2 {v_A}^2\over 3\delta(4-\delta^2)(4-\delta) w}\ ,
\end{equation}
where $w$ characterises the level of turbulence, and is equal to the
ratio of MHD wave energy density to magnetic field energy density.  The
free parameter in this relation is $v_A^2 /w$, where $v_A$ is the
Alfv\'en speed; we take $w = 1$ (\cite{SeoPtuskin94}).

\begin{figure}[t!]
   \centerline{
      \psfig{file=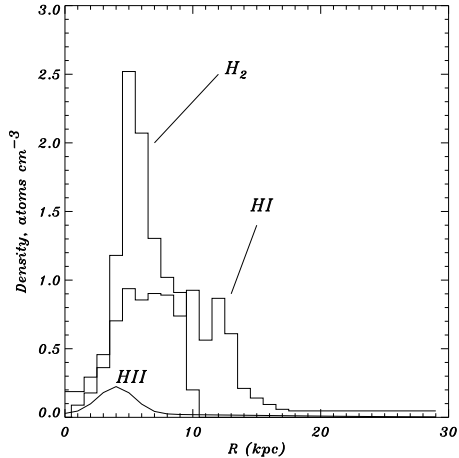,width=\fwb,clip=}
   }
\figcaption[fig1.ps]{The adopted radial distribution of atomic (HI),
molecular (H$_2$) and ionized (HII) hydrogen at $z = 0$.  
\label{fig1} }
\end{figure}

The adopted distributions of atomic and molecular hydrogen and of
ionized hydrogen are described in detail in \cite{SM98};
Fig.~\ref{fig1} shows the radial distribution of density in the
Galactic plane.  The He/H ratio of the interstellar gas is taken as
0.11 by number (see \cite{SM98} for a discussion).

\placefigure{fig1}

The distribution of cosmic-ray sources is chosen to reproduce (after
propagation) the cosmic-ray distribution determined by analysis of
EGRET \gray data (\cite{StrongMattox96}).  The form used is
\begin{equation}
\label{2.2}
q(R,z) = q_0 \left({R\over R_\odot}\right)^\eta e^{-\xi{R-R_\odot\over
R_\odot} -{|z|\over 0.2{\rm\ kpc}}}\ ,
\end{equation}
where $q_0$ is a normalization constant, $\eta$ and $\xi$ are
parameters; the $R$-dependence has the same parameterization as that
used for SNR by Case \& Bhattacharya (1996, 1998).  We compute models
with their SNR distribution, but also with different parameters to
better fit the \gray gradient.  We apply a cutoff in the source
distribution at $R = 20$ kpc since it is unlikely that significant
sources are present at such large radii.  The $z$-dependence of $q$ is
nominal and reflects simply the assumed confinement of sources to the
disk.

The primary propagation is computed first giving the primary
distribution as a function of ($R, z, p$); then the secondary source
function is obtained from the gas density and cross sections, and
finally the secondary propagation is computed.
The bremsstrahlung and inverse Compton \grays are computed
self-consistently from the gas and radiation fields used for the
propagation. The $\pi^0$-decay \grays are calculated explicitly
from the proton and Helium spectra using Dermer's (1986) approach. The
secondary nucleon and secondary $e^\pm$ source functions are computed
from the propagated primary distribution and the gas distribution.

\section{Propagation equation} \label{propagation_eq}

The propagation equation we use is written in the form:
\begin{equation}
\label{A.1}
{\partial \psi \over \partial t} 
= q(\vec r, p) 
+ \vec\nabla \cdot ( \Dxx\vec\nabla\psi - \vec V\psi )
+ \ddp\, p^2 \Dpp \ddp\, {1\over p^2}\, \psi
- {\partial\over\partial p} \left[\dot{p} \psi
- {p\over 3} \, (\vec\nabla \cdot \vec V )\psi\right]
- {1\over\tau_f}\psi - {1\over\tau_r}\psi\ ,
\end{equation}
where $\psi=\psi (\vec r,p,t)$ is the density per unit of total
particle momentum, $\psi(p)dp = 4\pi p^2 f(\vec p)$ in terms of
phase-space density $f(\vec p)$, $q(\vec r, p)$ is the source term,
$\Dxx$ is the spatial diffusion coefficient, $\vec V$ is the convection
velocity, reacceleration is described as diffusion in momentum space
and is determined by the coefficient $\Dpp$, $\dot{p}\equiv dp/dt$
is the momentum loss rate, $\tau_f$ is the time scale for
fragmentation, and $\tau_r$ is the time scale for the radioactive
decay. 

The numerical solution of the transport equation is based on a
Crank-Nicholson (\cite{Press92}) implicit second-order scheme.  The three spatial boundary conditions 
\begin{equation}
\label{B.4}
\psi(R_h,z,p) = \psi(R,\pm z_h,p) = 0
\end{equation}
are imposed on each iteration.

We use particle momentum as the kinematic variable since it greatly
facilitates the inclusion of the diffusive reacceleration terms.  The
injection spectrum of primary nucleons is assumed to be a power law in
momentum for the different species, $dq(p)/dp \propto p^{-\Gamma}$ for
the injected {\it density} as expected for diffusive shock acceleration
(e.g., \cite{Blandford80}). This corresponds to an injected {\it flux}
per kinetic energy interval $dF(E_k)/dE_k \propto p^{-\Gamma}$, a form
often used; the value of $\Gamma$ can vary with species.  The injection
spectrum for $^{12}$C and $^{16}$O was taken as $dq(p)/dp \propto
p^{-2.35}$, for the case of no reacceleration, and $p^{-2.25}$ with
reacceleration.  These values are consistent with Engelmann et
al.\ (1990) who give an injection index $2.23\pm0.05$. The same indices
reproduce the observed proton and $^4$He spectra
(\cite{MS98a}).  For primary electrons, the injection
spectrum can be adjusted to reproduce direct measurements or \gray and
synchrotron data; all details are given in our series of papers
(I--V).

For secondary nucleons, the source term is $q(\vec r, p) = \beta c\,
\psi_p (\vec r, p)[\sigma^{ps}_H (p) n_H (\vec r)+ \sigma^{ps}_{He}(p)
n_{He}(\vec r)]$, where $\sigma^{ps}_H (p)$, $\sigma^{ps}_{He} (p)$ are
the production cross sections for the secondary from the progenitor on
H and He targets, $\psi_p$ is the progenitor density, and $n_H$,
$n_{He}$ are the interstellar hydrogen and Helium number densities.

To compute \BC\ and \Berat\ it is sufficient for our purposes to treat
only one principal progenitor and compute weighted cross sections based
on the observed cosmic-ray abundances, which we took from Lukasiak et
al. (1994b).  Explicitly, for a principal primary with abundance $I_p$,
we use for the production cross section $\overline\sigma^{ps} = \sum_i
\sigma^{is} I_i/I_p$, where $\sigma^{is}$, $I_i$ are the cross sections
and abundances of all species producing the given secondary.  For the
case of Boron, the Nitrogen progenitor is secondary but only accounts
for $\approx$ 10\% of the total Boron production, so that the
approximation of weighted cross sections is sufficient.

For the fragmentation cross sections we use the formula given by Letaw,
Silberberg, \& Tsao (1983).  For the secondary production cross sections
we use the Webber, Kish, \& Schrier (1990) parameterizations in the form
of code obtained from the Transport Collaboration (\cite{Guzik97}).
For the important \BC\ ratio, we take the $^{12}$C, $^{16}$O $\to\,
^{10}$B, $^{10}$C, $^{11}$B, $^{11}$C cross sections from the fit to
experimental data given by Heinbach \& Simon (1995).  For electrons and
positrons the same propagation equation is valid when the appropriate
energy loss terms (ionization, bremsstrahlung, inverse Compton,
synchrotron) are used. The energy loss formulae for these loss
mechanisms are given in \cite{SM98}.

\section{Evaluation of models}

We consider the cases of diffusion+convection and
diffusion+reacceleration, since these are the minimum combinations
which can reproduce the key observations. In principle all three
processes could be significant, and such a general model can be
considered if independent astrophysical information or models, for
example for a Galactic wind (e.g., \cite{Zirakashvili96,Ptuskin97}),
were to be used.

In our evaluations we use the \BC\ data summarized by Webber et
al.\ (1996), from HEAO--3 and Voyager 1 and 2.  The spectra were
modulated to 500 MV appropriate to this data using the force-field
approximation (\cite{GleesonAxford68}).  We also show \BC\ values from
Ulysses (\cite{DuVernois96}) for comparison, but since this has large
modulation (600--1080 MV) we do not base conclusions on these
values.  We use the measured \Berat\ ratio from Ulysses
(\cite{Connell98}) and from Voyager--1,2, IMP--7/8, ISEE--3 as
summarized by Lukasiak et al. (1994a).

The source distribution adopted has $\eta=0.5$, $\xi=1.0$ in
eq.~(\ref{2.2}) (apart from the cases with SNR source distribution).
This form adequately reproduces the small observed \gray based
gradient, for all $z_h$; a more detailed discussion is given in 
Section~\ref{CRgradients}.

\subsection{Diffusion/convection models}

The main parameters are $z_h$, $D_0$, $\delta_1$, $\delta_2$ and
$\rho_0$ and $dV/dz$.  We treat $z_h$ as the main unknown quantity, and
consider values 1--20 kpc.  For a given $z_h$ we show \BC\ for a series
of models with different $dV/dz$.

\begin{figure}[t!]
   \centerline{
      \psfig{file=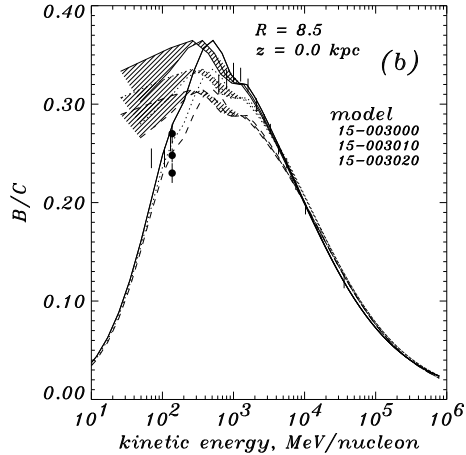,width=\fwb,clip=}
   }
\figcaption[fig2.ps]{ \BC\ ratio for
diffusion/convection models without break in diffusion coefficient, for
$z_h$ = 3 kpc, $dV/dz$ = 0 (solid line), 5 (dotted line), and 10 km
s$^{-1}$ kpc$^{-1}$ (dashed line).  Solid line: interstellar ratio,
shaded area: modulated to 300--500 MV.  Data: vertical bars:  HEAO-3,
Voyager (\cite{Webber96}), filled circles: Ulysses (\cite{DuVernois96}:
$\Phi$ = 600, 840, 1080 MV).
\label{fig2} }
\end{figure}

\placefigure{fig2}

Fig.~\ref{fig2} shows the case of no break, $\delta_1 = \delta_2$;
for each $dV/dz$, the remaining parameters $D_0$, $\delta_1$ and
$\rho_0$ are adjusted to fit the data as well as possible.  It is clear
that a {\it good} fit is {\it not} possible; the basic effect of
convection is to reduce the variation of \BC\ with energy, and although
this improves the fit at low energies the characteristic peaked shape
of the measured \BC\ cannot be reproduced.  Although modulation makes
the comparison with the low energy Voyager data somewhat uncertain,
Fig.~\ref{fig2} shows that the fit is unsatisfactory; the same is
true even if we use a very low modulation parameter (300 MV) in an
attempt to improve the fit.  This modulation is near the minimum value
for the entire Voyager 17 year period (cf. the average value of 500 MV;
\cite{Webber96}).  The failure to obtain a good fit is an important
conclusion since it shows that the simple inclusion of convection
cannot solve the problem of the low-energy falloff in \BC.

\begin{figure}[t!]
   \centerline{
      \psfig{file=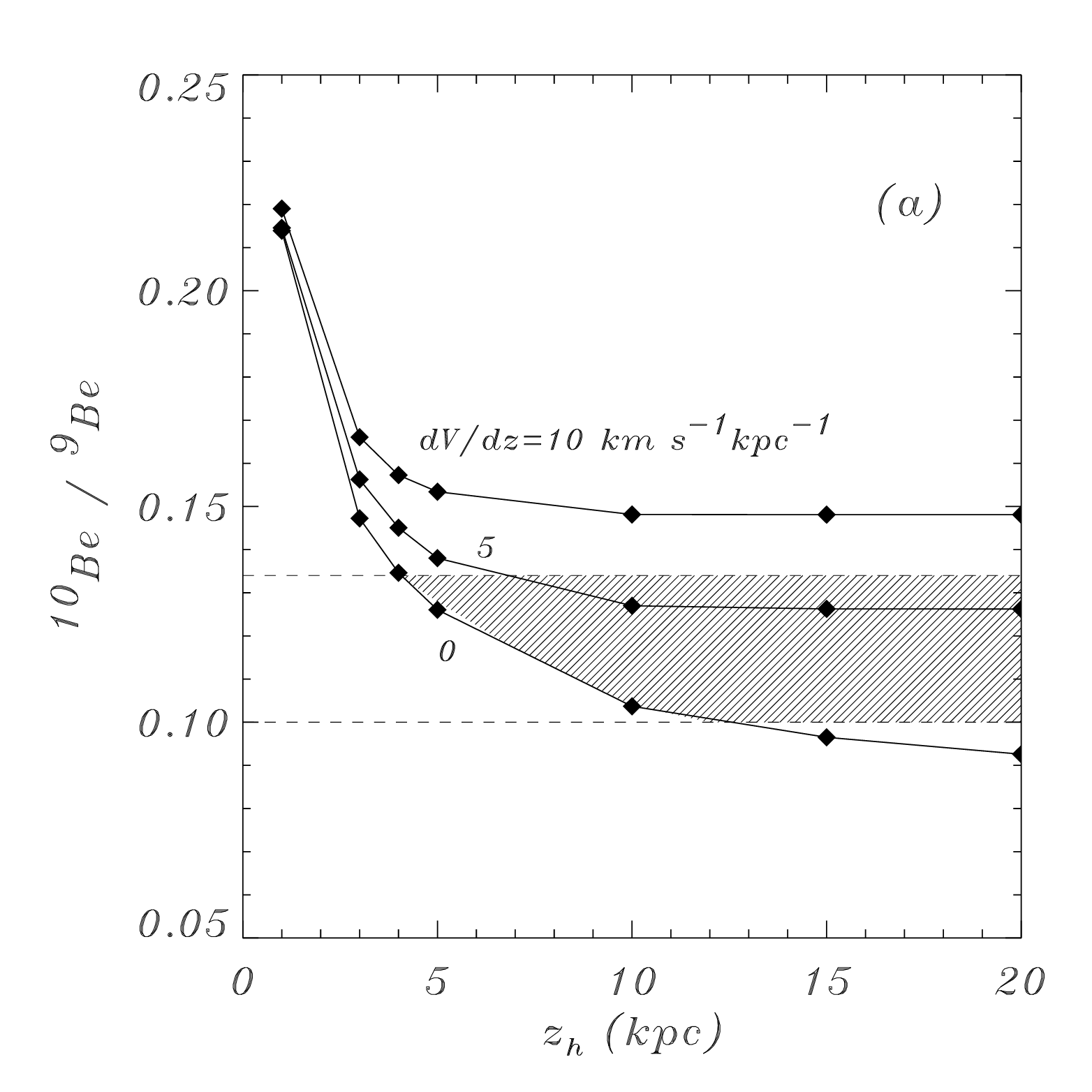,width=\fwb,clip=}
      \psfig{file=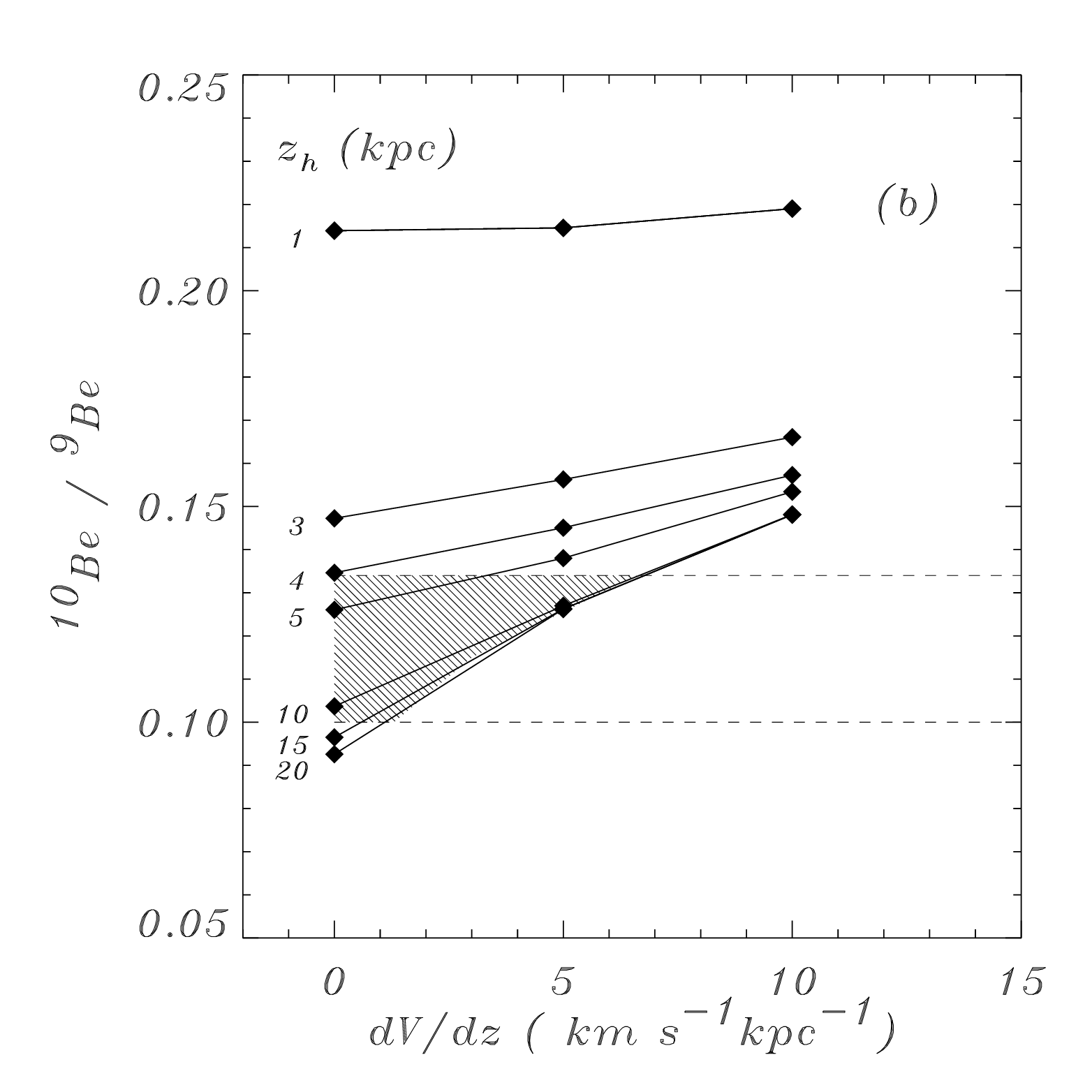,width=\fwb,clip=}
   }
\figcaption[fig3a.ps,fig3b.ps]{ Predicted \Berat\ ratio as function of
(a) $z_h$ for $dV/dz$ = 0, 5, 10 km s$^{-1}$ kpc$^{-1}$, (b) $dV/dz$
for $z_h = 1 - 20$ kpc at 525 MeV/nucleon corresponding to the mean
interstellar value for the Ulysses data (\cite{Connell98}); the Ulysses
experimental limits are shown as horizontal dashed lines.  The shaded
regions show the parameter ranges allowed by the data.
\label{fig3} }
\end{figure}

\placefigure{fig3}

We can however force a fit to the data by allowing a break in
$\Dxx(p)$, i.e.\ $\delta_1 \ne \delta_2$.  In the absence of
convection, the falloff in \BC\ at low energies requires that the
diffusion coefficient increases rapidly below $\rho_0 = 3$ GV
($\delta_1\sim -0.6$) reversing the trend from higher energies
($\delta_2 \sim +0.6$).  Inclusion of the convective term does not
reduce the size of the {\it ad hoc} break in the diffusion coefficient,
in fact it rather exacerbates the problem by requiring a larger
break\footnote{  Note that the dependence of interaction rate on
particle velocity itself is not sufficient to cause the full observed
low-energy falloff.  In leaky-box treatments the low-energy behaviour
is modelled by adopting a constant path-length below a few GeV/nucleon,
without attempting to justify this physically.  A convective term is
often invoked, but our treatment shows that this alone is not
sufficient.  }.

Fig.~\ref{fig3} summarizes the limits on $z_h$ and $dV/dz$, using the
\Berat\ ratio at the interstellar energy of 525 MeV/nucleon appropriate
to the Ulysses data (\cite{Connell98}).  For $z_h <4$ kpc, the
predicted ratio is always too high, even for no convection; no
convection is allowed for such $z_h$ values since this increases
\Berat\ still further.  For $z_h \ge 4$ kpc agreement with \Berat\ is
possible provided $0 < dV/dz < 7$ km s$^{-1}$ kpc$^{-1}$.  We
conclude from Fig.~\ref{fig3}a that in the absence of convection
$4{\rm\ kpc}<z_h < 12 {\rm\ kpc}$, and if convection is allowed the
lower limit remains but no upper limit can be set.  It is interesting
that an upper as well as a lower limit on $z_h$ is obtained in the case
of no convection, although \Berat\ approaches asymptotically a constant
value for large halo sizes and becomes insensitive to the halo
dimension. From Fig.~\ref{fig3}b, $dV/dz < 7$ km s$^{-1}$
kpc$^{-1}$ and this figure places upper limits on the convection
parameter for each halo size. These limits are rather strict, and a
finite wind velocity is only allowed in any case for $z_h > 4$ kpc.
Note that these results are not very sensitive to modulation since the
predicted \Berat\ is fairly constant from 100 to 1000 MeV/nucleon.

\subsection{Diffusive reacceleration models \label{diff_reacc_models}}

\begin{figure}[t!]
   \centerline{
      \psfig{file=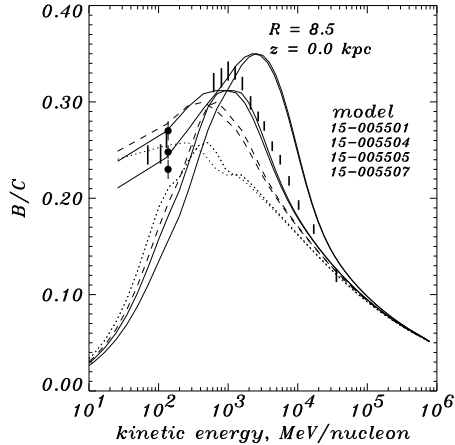,width=\fwb,clip=}
   }
\figcaption[fig4.ps]{ \BC\ ratio for diffusive reacceleration models
with $z_h$ = 5 kpc, $v_A$ = 0 (dotted), 15 (dashed), 20 (thin solid), 30 km
s$^{-1}$ (thick solid).  
In each case the interstellar ratio and the ratio modulated to 500 MV
is shown.  Data: as Fig.~\ref{fig2}. 
\label{fig4} }
\end{figure}

\placefigure{fig4}

\begin{figure}[t!]
   \centerline{
      \psfig{file=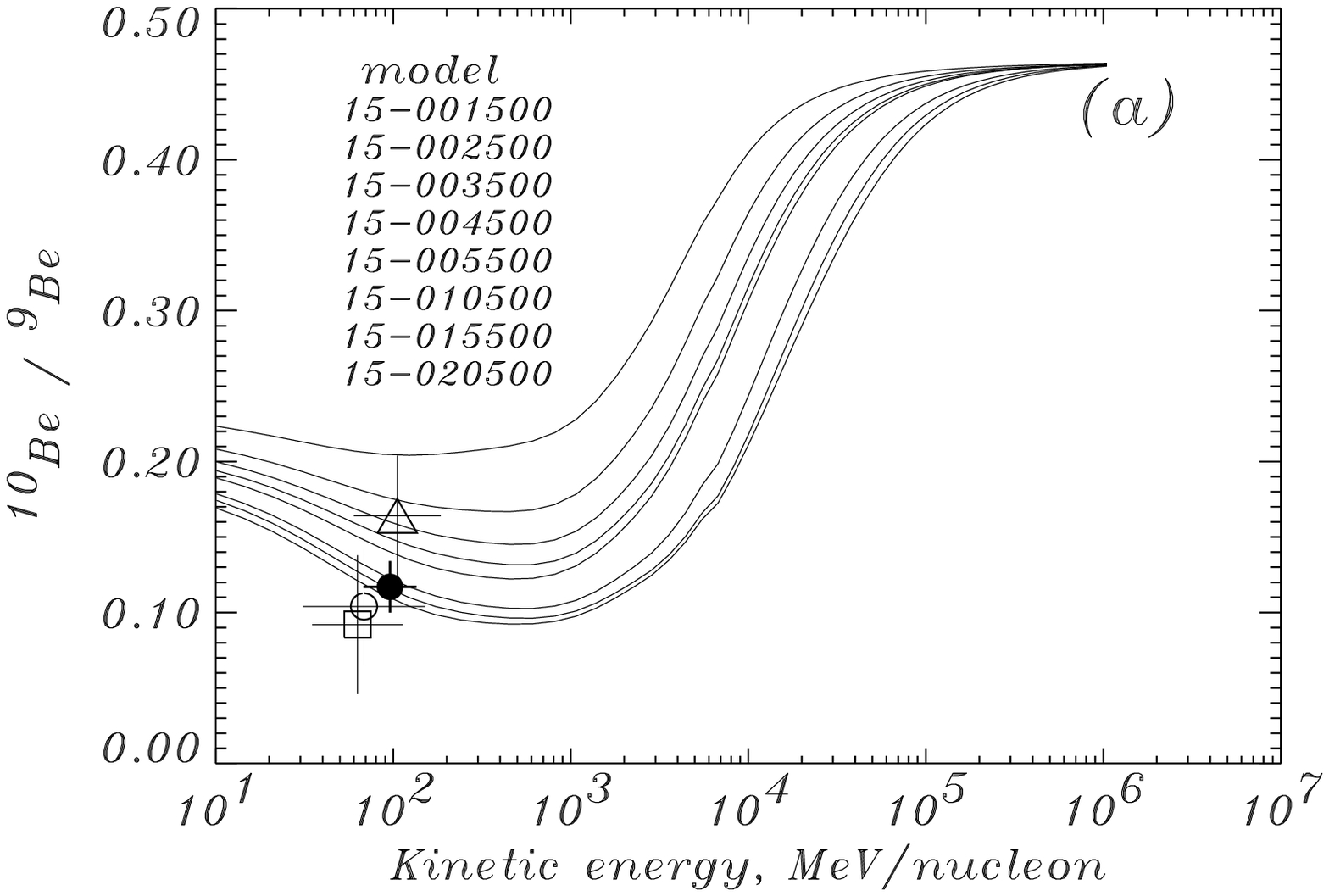,width=\fwb,clip=}
      \psfig{file=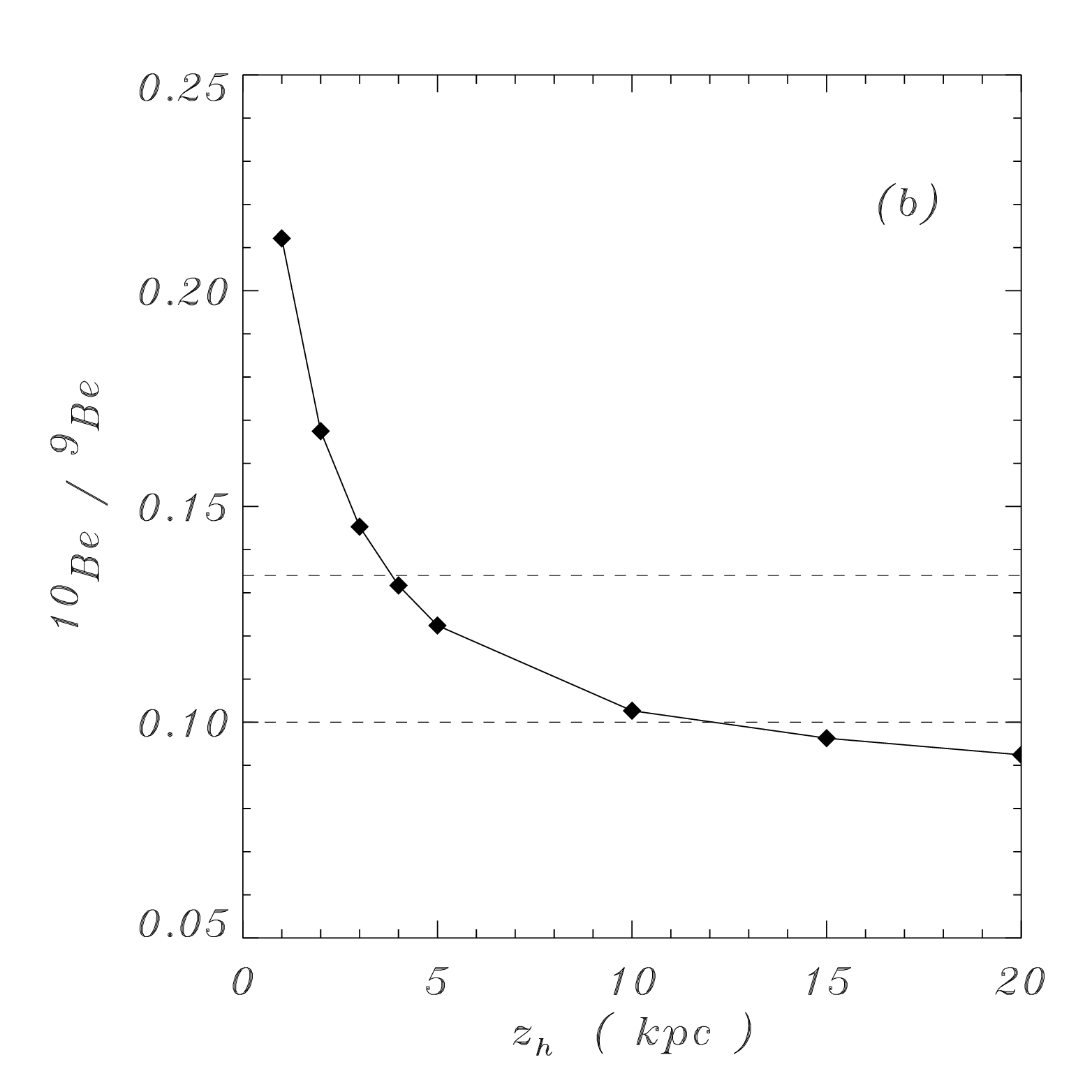,width=\fwb,clip=}
   }
\figcaption[fig5a.ps,fig5b.ps]{ \Berat\ ratio for diffusive
reacceleration models: (a) as function of energy for (from top to
bottom) $z_h$ = 1, 2, 3, 4, 5, 10, 15 and 20 kpc; (b) as function of
$z_h$ at 525 MeV/nucleon corresponding to the mean interstellar value
for the Ulysses data (\cite{Connell98}); the Ulysses experimental
limits are shown as horizontal dashed lines.  Data points from Lukasiak
et al. (1994a) (Voyager-1,2: square, IMP-7/8: open circle, ISEE-3:
triangle) and Connell (1998) (Ulysses): filled circle.
\label{fig5} }
\end{figure}

\placefigure{fig5}

The main parameters are $z_h$, $D_0$ and $v_A$.  Again we treat $z_h$
as the main unknown quantity.  The evaluation is simpler than for
convection models since the number of free parameters is smaller.
Fig.~\ref{fig4} illustrates the effect on \BC\ of varying $v_A$, from
$v_A = 0$ (no reacceleration) to $v_A=30$ km s$^{-1}$, for $z_h= 5$
kpc.  This shows how the initial form becomes modified to produce the
characteristic peaked shape.  Reacceleration models thus lead naturally
to the observed peaked form of \BC, as pointed out by several previous
authors (e.g., \cite{Letaw93,SeoPtuskin94,HeinbachSimon95}).

Fig.~\ref{fig5} shows \Berat\ for the same models,
(a) as a function of energy for various $z_h$, (b) as a function of
$z_h$ at 525 MeV/nucleon corresponding to the Ulysses measurement.
Comparing with the Ulysses data point, we conclude that $4{\rm\ kpc}
<z_h < 12$ kpc. Again the result is not very sensitive to modulation
since the predicted \Berat\ is fairly constant from 100 to 1000
MeV/nucleon. 

Energy losses attenuate the flux of stable nuclei much more than
radioactive nuclei, and hence lead to an increase in \Berat.  Clearly
if losses are ignored the predicted ratio will be too low and the
derived value of $z_h$ will be too small since $z_h$ will have to be
reduced to fit the observations. 

Our results on the halo size can be compared with those of other
studies: $z_h \ge7.8$ kpc (\cite{Freedman80}), $ z_h \le3$ kpc
(\cite{Bloemen93}), and $z_h \le4$ kpc (\cite{Webber92}). 
Lukasiak et al. (1994a) found $1.9{\rm\ kpc} < z_h < 3.6$ kpc (for no
convection) based on Voyager Be data and using the Webber, Lee, \& Gupta
(1992) models.  We believe our new limits to be an improvement, first
because of the improved Be data from Ulysses, second because of our
treatment of energy losses (see Section~\ref{diff_reacc_models}) and
generally more realistic astrophysical details in our model.  
Recently, Webber \& Soutoul (1998), Ptuskin \& Soutoul (1998) have
obtained $z_h= 2-4$ kpc and $4.9_{-2}^{+4}$ kpc, respectively, in 
agreement with our results.

\section{Cosmic-ray gradients} \label{CRgradients}

\begin{figure}[t!]
   \centerline{
      \psfig{file=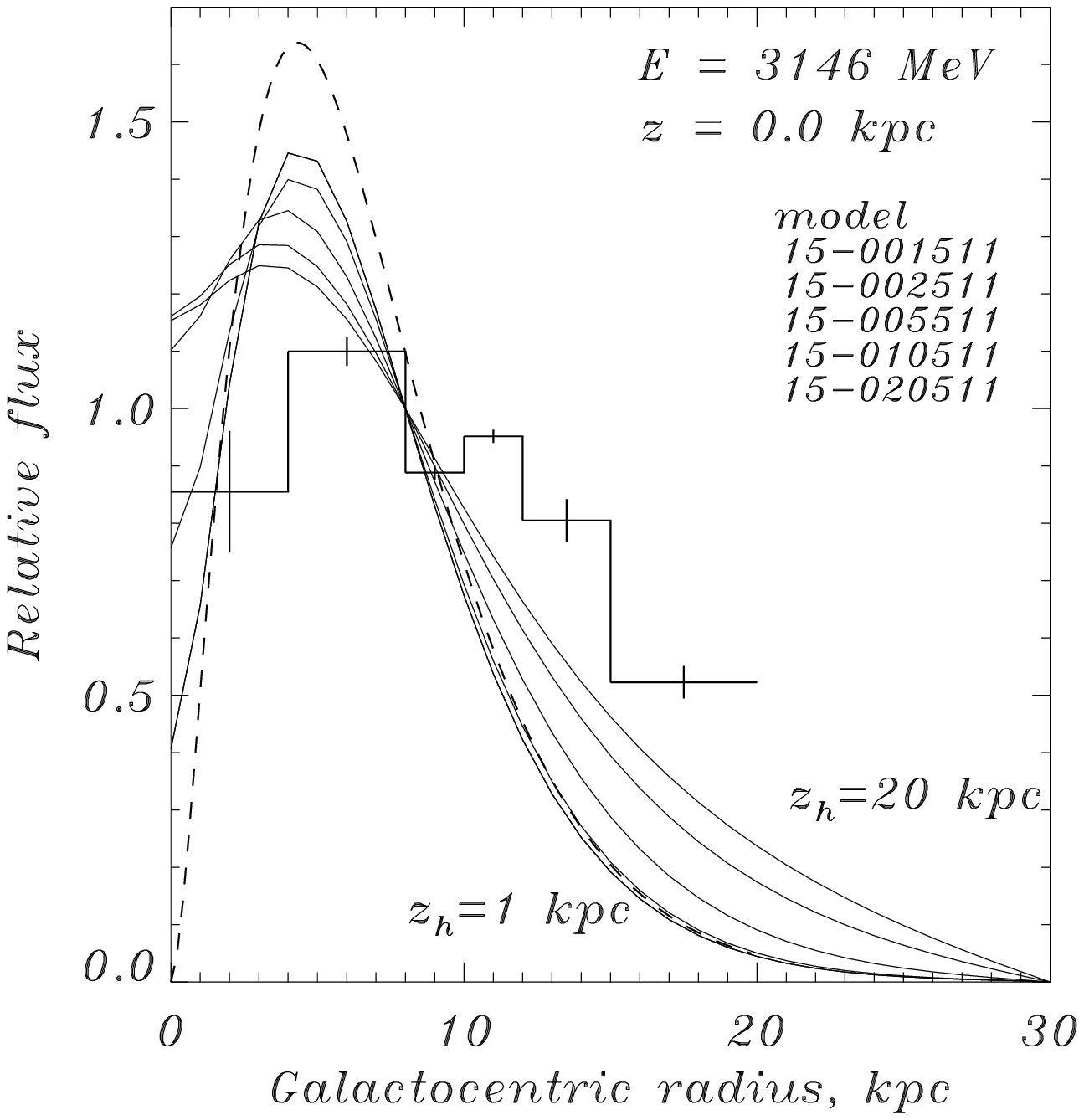,width=\fwb,clip=}
      \psfig{file=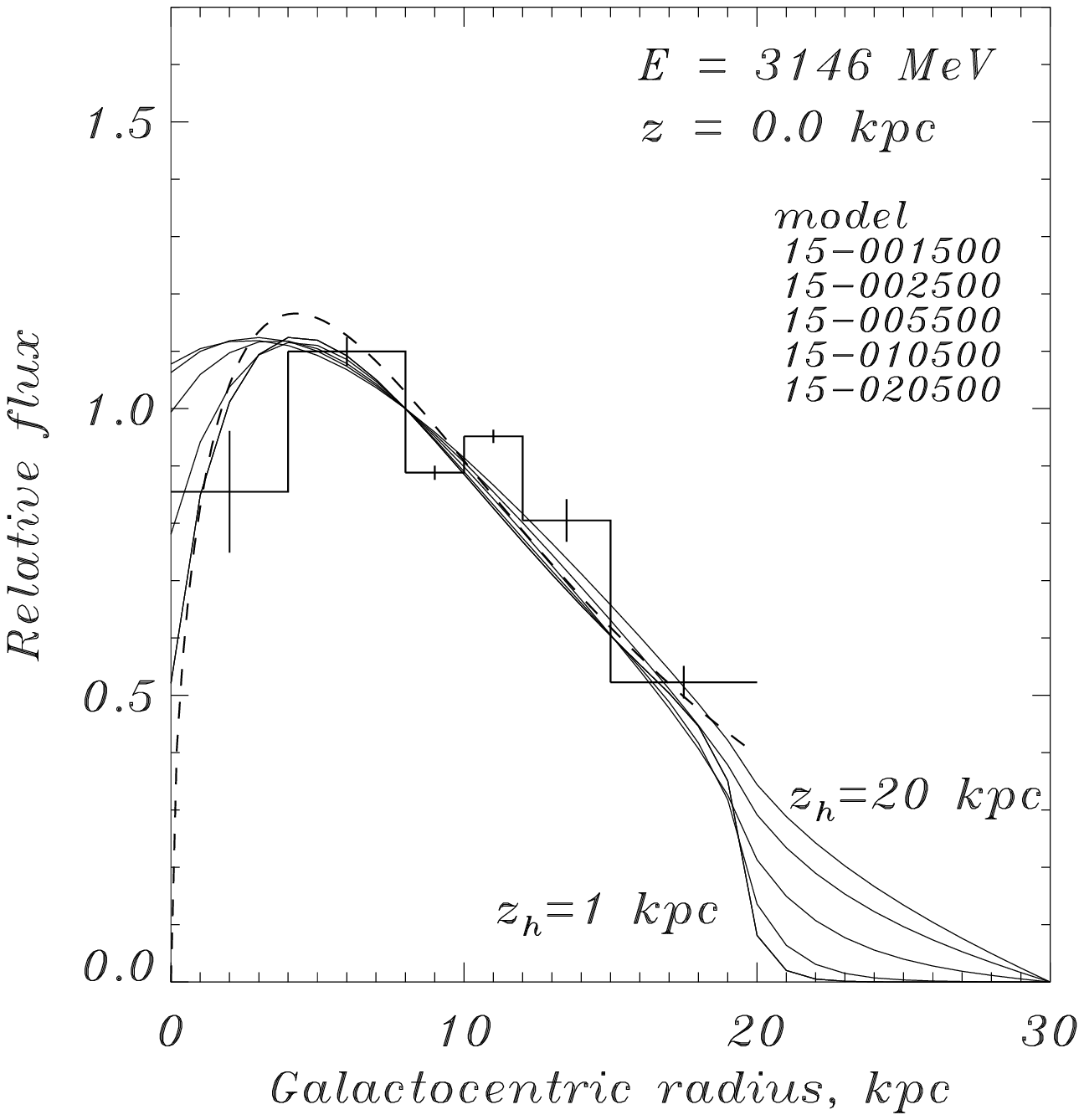,width=\fwb,clip=}
   }
\figcaption[fig6a.ps,fig6b.ps]{ {\it Left panel}:
Radial distribution of 3 GeV protons at $z = 0$,
for diffusive reacceleration model with halo sizes $z_h = 1$, 3, 5,
10, 15, and 20 kpc (solid curves). The source distribution is that for
SNR given by Case \& Bhattacharya (1996), shown as a dashed line. The
cosmic-ray distribution deduced from EGRET $>$100 MeV \grays
(\cite{StrongMattox96}) is shown as the histogram.  
{\it Right panel}:
Radial distribution of 3 GeV protons at $z = 0$,
for diffusive reacceleration model with various halo sizes $z_h = 1$,
3, 5, 10, 15, and 20 kpc (solid curves). The source distribution used
is shown as a dashed line. It was adopted to reproduce the
cosmic-ray distribution deduced from EGRET $>$100 MeV \grays
(\cite{StrongMattox96}) which is shown as the histogram.
\label{fig6} }
\end{figure}

\placefigure{fig6}

An important constraint on any model of cosmic-ray propagation is
provided by \gray data which give information on the radial
distribution of cosmic rays in the Galaxy. For a given source
distribution, a large halo will give a smaller cosmic-ray gradient.  It
is generally believed that supernova remnants (SNR) are the main
sources of cosmic rays (see \cite{Webber97} for a recent review), but
unfortunately the distribution of SNR is poorly known due to selection
effects.  Nevertheless it is interesting to compare quantitatively the
effects of halo size on the gradient for a plausible SNR source
distribution.  For illustration we use the SNR distribution from Case
\& Bhattacharya (1996), which is peaked at $R = 4 - 5$ kpc and has a
steep falloff towards larger $R$.

Fig.~\ref{fig6} (left panel) shows the effect of halo size on the
resulting radial distribution of 3 GeV cosmic-ray protons, for the
reacceleration model.  For comparison we show the cosmic-ray
distribution deduced by model-fitting to EGRET \gray data ($>100$ MeV)
from Strong \& Mattox (1996), which is dominated by the $\pi^0$-decay
component generated by GeV nucleons; the analysis by Hunter et al.
(1997), based on a different approach, gives a similar result.  The
predicted cosmic-ray distribution using the SNR source function is too
steep even for large halo sizes; in fact the halo size has a relatively
small effect on the distribution.  Other related distributions such as
pulsars (\cite{Taylor93}, \cite{Johnston94}) have an even steeper
falloff.  Only for $z_h = 20$ kpc does the gradient approach that
observed, and in this case the combination of a large halo and a
slightly less steep SNR distribution could give a satisfactory fit.
For diffusion/convection models the situation is similar, with more
convection tending to make the gradient follow more closely the
sources.  A larger halo ($z_h \gg 20$ kpc), apart from being excluded
by the $^{10}$Be analysis presented here, would in fact not improve the
situation much since Fig.~\ref{fig6} shows that the gradient
approaches an asymptotic shape which hardly changes beyond a certain
halo size.  This is a consequence of the nature of the diffusive
process, which even for an unlimited propagation region still retains
the signature of the source distribution.

Based on these results we have to conclude, in the context of the
present models, that the distribution of sources is not that expected
from the (highly uncertain: see \cite{Green91}) distribution of SNR.
This conclusion is similar to that previously found by others
(\cite{Webber92,Bloemen93}).  In view of the difficulty of deriving the
SNR distribution this is perhaps not a serious shortcoming; if SNR are
indeed cosmic-ray sources then it is possible that the \gray analysis
gives the best estimate of their Galactic distribution.  Therefore in
our standard model we have obtained the source distribution empirically
by requiring consistency with the high energy \gray results.

Alternatively it is possible that the diffusion is not isotropic but
occurs preferentially in the radial direction, so smoothing the source
distribution more effectively. This possibility will be
addressed in future work.

Fig.~\ref{fig6} (right panel) shows the source distribution adopted
in the present work, and the resulting 3 GeV proton distribution, again
compared to that deduced from $\gamma$-rays. The gradients are now
consistent, especially considering that some systematic effects, due
for example unresolved \gray sources, are present in the \gray
based distribution.

\section{Interstellar positrons and antiproton spectra}

The positron and antiproton fluxes reflect the proton and Helium
spectra throughout the Galaxy and thus provide an essential check on
propagation models and also on the interpretation of diffuse \gray
emission (\cite{MSR98}).  Secondary positrons and antiprotrons in
Galactic cosmic rays are produced in collisions of cosmic-ray particles
with interstellar matter\footnote{  Secondary origin of cosmic-ray
antiprotons is basically accepted, though some other exotic
contributors such as, e.g., neutralino annihilation (\cite{Bottino98})
are also discussed.  }.  These are an important diagnostic for models of
cosmic-ray propagation and provide information complementary to that
provided by secondary nuclei.  However, unlike secondary nuclei,
antiprotons reflect primarily the propagation history of the protons,
the main cosmic-ray component.

In our model the proton and Helium spectra are computed as a function
of $(R,z,p)$ by the propagation code.  The injection spectrum is
adjusted to give a good fit to the locally measured spectrum,
normalizing at 10 GeV/nucleon.

For the injection spectra of protons, we find $\Gamma=2.15$ reproduces
the observed spectra in the case of no reacceleration, and
$\Gamma=2.25$ with reacceleration.  We find it is necessary to use
slightly steeper (0.2 in the index) injection spectra for Helium nuclei
in order to fit the observed spectra in the 1--100 GeV range of
interest for positron production.  The spectra fit up to about 100 GeV
beyond which the Helium spectrum without reacceleration becomes too
steep and the proton spectrum with reacceleration too flat; these
deviations are of no consequence for the positron and antiproton 
calculations.

\begin{figure}[t!]
   \centerline{
      \psfig{file=fig7a.ps,width=\fwb,clip=}
      \psfig{file=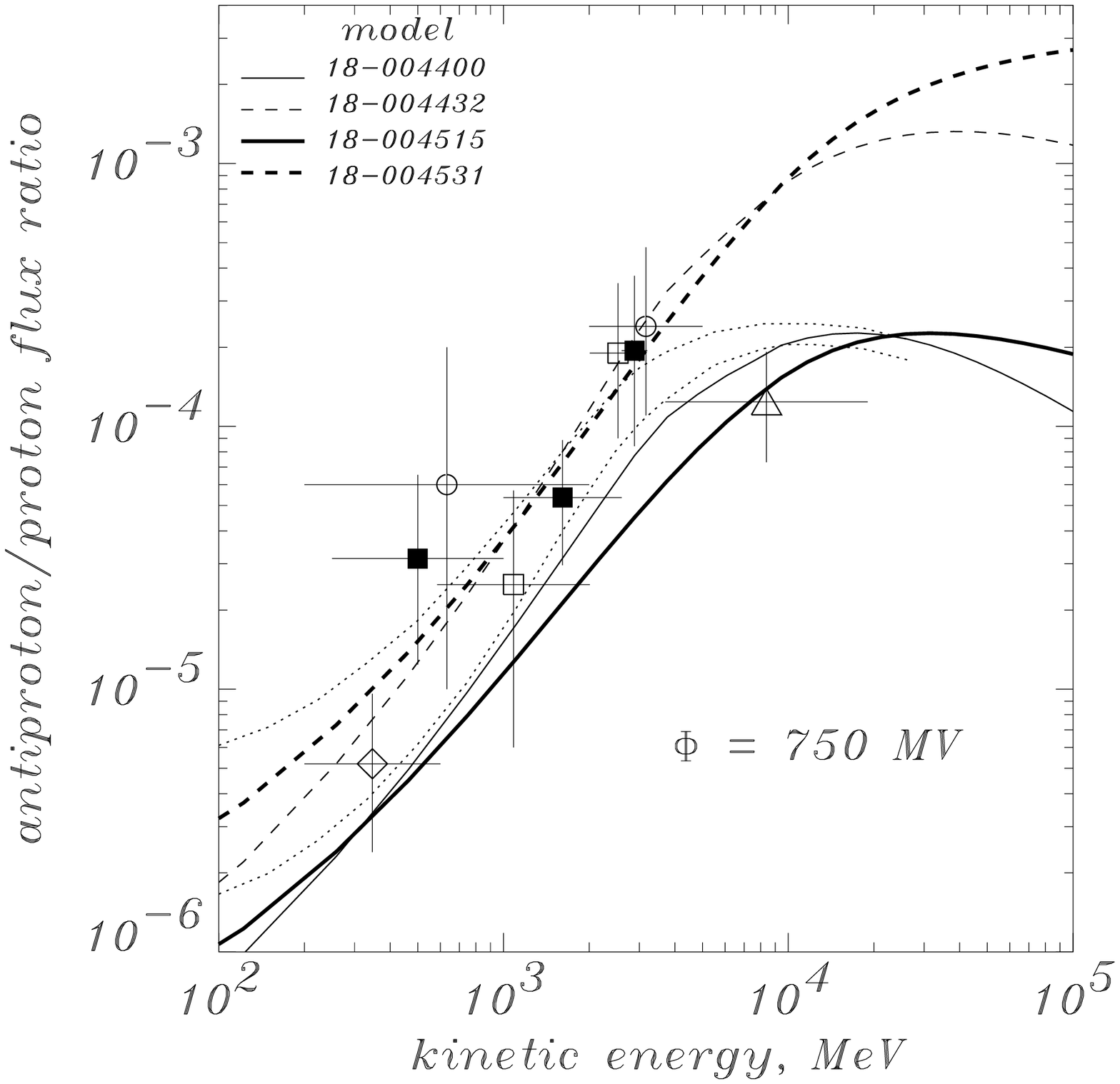,width=\fwb,clip=}
   }
   \caption[fig7a.ps,fig7b.ps]{
{\it Left panel:} Interstellar nucleon and antiproton spectra as
calculated in nonreacceleration models (thin lines) and models with
reacceleration (thick lines). Proton spectra consistent with the local
one are shown by the solid lines, hard spectra are shown by the dashed
lines.  The local spectrum as measured by IMAX (\cite{Menn97}) is shown
by dots.
{\it Right panel:} $\bar{p}/p$ ratio for different ambient proton
spectra. Lines are coded as on the left. The ratio is modulated with
$\Phi=750$ MV.  Calculations of Simon et al.\ (1998) are
shown by the dotted lines. Data: see references in \cite{MSR98}.
\label{fig7}}
\end{figure}

Our calculations of the interstellar antiproton spectra and $\bar{p}/p$
ratio for these spectra are shown in Fig.~\ref{fig7}.  The computed
antiproton spectrum is divided by the same interstellar proton
spectrum, and the ratio is modulated to 750 MV.  The corresponding
ratios are shown on the right panel. We have performed the same
calculations for models with and without reacceleration and the results
differ only in details.  As seen, our result agrees well with the
calculations of Simon et al.\ (1998), showing that our treatment of the
production cross-sections is adequate (for the details of the cross
sections see \cite{MSR98}).

\placefigure{fig7}

\begin{figure}[t!]
   \centerline{
      \psfig{file=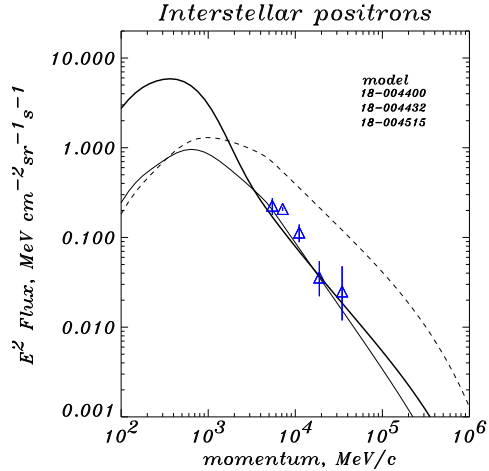,width=\fwb,clip=}
   }
\figcaption[fig8.ps]{
Spectra of secondary positrons for `conventional' (thin
line) and hard (dashes) nucleon spectra (no reacceleration).  Thick
line: `conventional' case with reacceleration.  Data: Barwick et
al.\ (1998).
\label{fig8}}
\end{figure}

\placefigure{fig8}

Fig.~\ref{fig8} shows the computed secondary positron spectra
for the cases without and with reacceleration.  Our predictions are
compared with recent absolute measurements above a few GeV where solar
modulation is small (\cite{Barwick98}), and the agreement is
satisfactory in both cases; this comparison has
the advantage of being independent of the electron spectrum, unlike the
positron/electron ratio which was the main focus of
\cite{MS98a}.

\section{Probes of the interstellar nucleon spectrum} \label{positrons_pbar}

Diffuse Galactic \gray observations have been interpreted as requiring
a harder average nucleon spectrum in interstellar space than that
observed directly (\cite{Hunter97}, \cite{Gralewicz97}, \cite{Mori97},
\cite{MS98b},c, see also Section \ref{gammarays}).  A
sensitive test of the interstellar nucleon spectra is provided by
secondary antiprotons and positrons.  Because they are secondary, they
reflect the {\it large-scale} nucleon spectrum independent of local
irregularities in the primaries.

\begin{figure}[t!]
   \centerline{
      \psfig{file=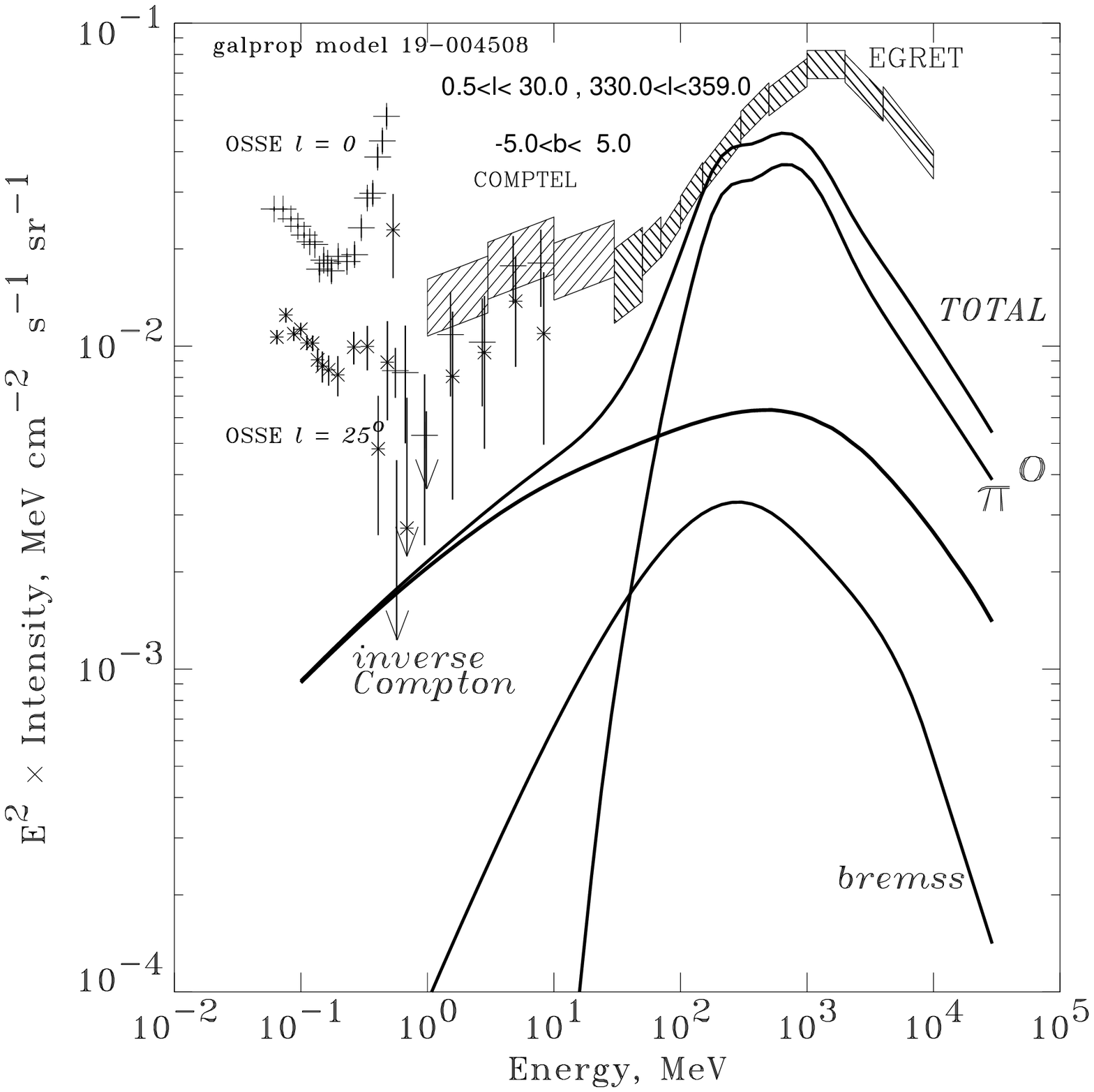,width=\fwb,clip=}
      \psfig{file=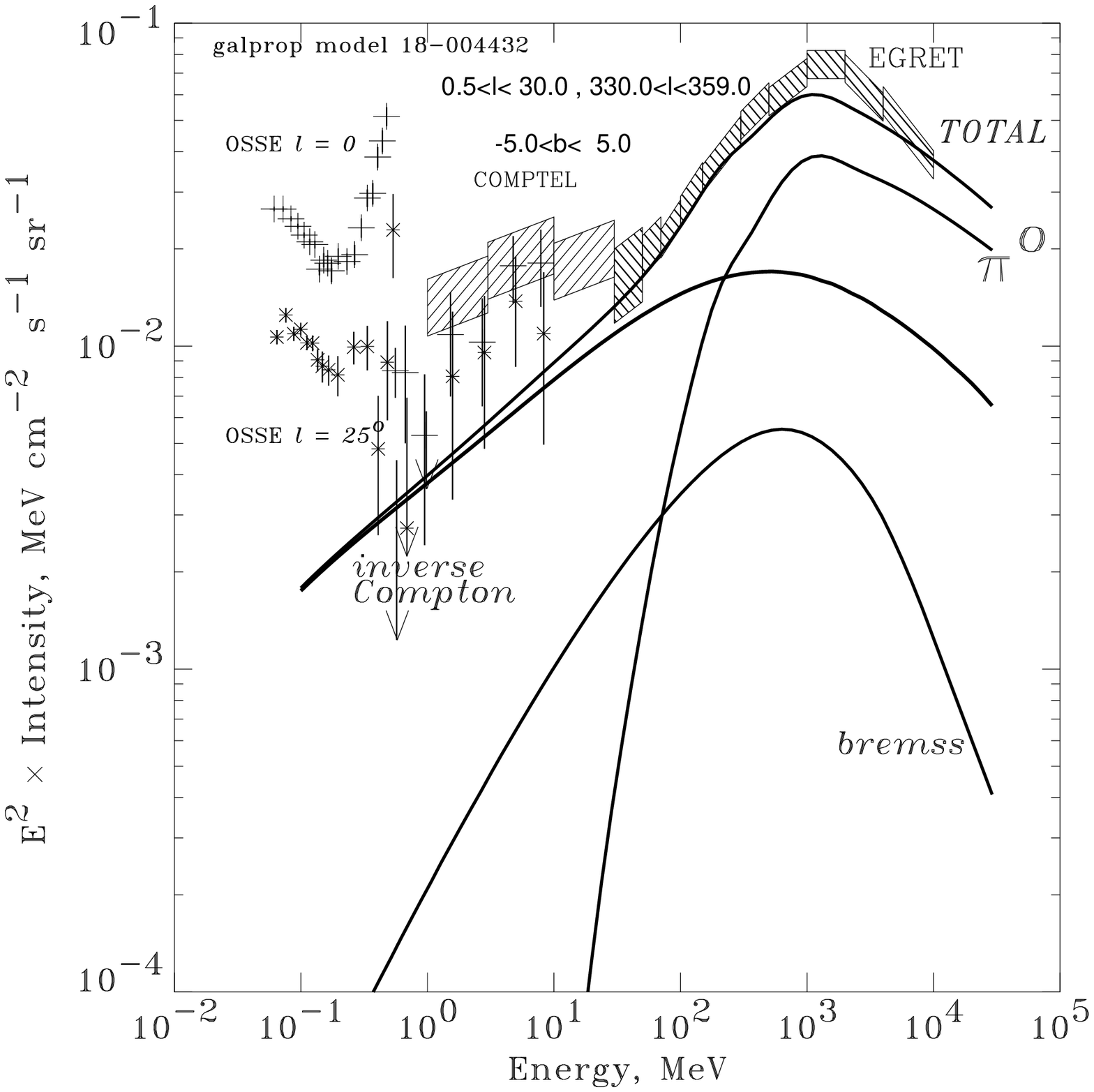,width=\fwb,clip=}
   }
   \caption[fig9a.ps,fig9b.ps]{
Gamma-ray energy spectrum of the inner Galaxy ($300^\circ \le l\le
30^\circ$, $|b|\le 5^\circ$) compared with our model calculations.
Data: EGRET (\cite{StrongMattox96}), COMPTEL (\cite{Strongetal98}),
OSSE ($l=0, 25^\circ$:  \cite{Kinzer97}).  {\it Left panel:} Model with
`conventional' nucleon and electron spectra.  Also shown are the
contributions of individual components:  bremsstrahlung, inverse
Compton, and $\pi^0$-decay.  {\it Right panel:} The same compared to
the model with the {\it hard nucleon} spectrum (no reacceleration).
\label{fig9}}
\end{figure}

We consider a case which matches the \gray data (Fig.~\ref{fig9}) at
the cost of a much harder proton spectrum than observed.  The dashed
lines in Fig.~\ref{fig7} (right) show the $\bar{p}/p$ ratio for the
hard proton spectrum (with and without reacceleration); the ratio is
still consistent with the data at low energies but rapidly increases
toward higher energies and becomes $\sim$4 times higher at 10 GeV.  Up
to 3 GeV it does not confict with the data with their very large error
bars.  It is however larger than the point at 3.7--19 GeV
(\cite{Hof96}) by about $5\sigma$. Clearly we cannot conclude
definitively on the basis of this one point\footnote{  We do not
consider here the older $\bar{p}$ measurement of Golden et al.\ (1984a)
because the flight of the early instrument in 1979 was repeated in 1991
with significantly improved instrument and analysis techniques (see
\cite{Hof96} and a discussion therein).  }, but it does indicate the
sensitivity of this test. In view of the sharply rising ratio in the
hard-spectrum scenario it seems unlikely that the data could be fitted
in this case even with some re-scaling due to propagation
uncertainties.  More experiments are planned (see \cite{MSR98} for a
summary) and these should allow us to set stricter limits on the
nucleon spectra including less extreme cases than considered here, and
to constrain better the interpretation of $\gamma$-rays.

\placefigure{fig9}

Positrons also provide a good probe of the nucleon spectrum, but are
more affected by energy losses and propagation uncertainties.
Fig.~\ref{fig8} shows, in addition to the normal case, the positron
flux resulting from a hard nucleon spectrum. The predicted flux is a
factor 4 above the Barwick et al. (1998) measurements and hence
provides further evidence against the `hard nucleon spectrum' 
hypothesis.

\section{Diffuse Galactic continuum gamma rays} \label{gammarays}

\begin{figure}[t!]
   \centerline{
      \psfig{file=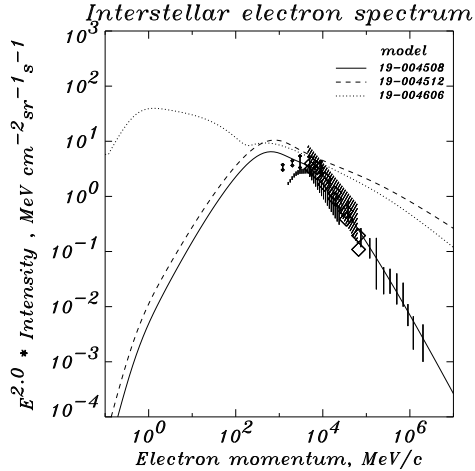,width=\fwb,clip=}
   }
\figcaption[fig10.ps]{
Electron spectra at $R_\odot = 8.5$ kpc in the plane, for
`conventional' (solid line), and hard electron spectrum models without
(dashes), and with (dots) low-energy upturn.  Data (direct
measurements):  Taira et al.\ (1993) (vertical lines), Golden et
al.\ (1984b, 1994) (shaded areas), Ferrando et al.\ (1996) (small
diamonds), Barwick et al.\ (1998) (large diamonds).
\label{fig10}}
\end{figure}

\placefigure{fig10}

We can also use our model to study the diffuse \gray emission from the
Galaxy.  Recent results from both COMPTEL and EGRET indicate that
inverse Compton (IC) scattering is a more important contributor to the
diffuse emission that previously believed.  COMPTEL results
(\cite{Strongetal97}) for the 1--30 MeV range show a latitude
distribution in the inner Galaxy which is broader than that of HI and
H$_2$, so that bremsstrahlung of electrons on the gas does not appear
adequate and a more extended component such as IC is required.  The
broad distribution is the result of the large $z$-extent of the
interstellar radiation field\footnote{  We have made a new calculation
of the interstellar radiation field (\cite{SMR98}) based on stellar
population models and IRAS and COBE data.  } which can interact with
cosmic-ray electrons up to several kpc from the plane.  At much higher
energies, the puzzling excess in the EGRET data above 1 GeV relative to
that expected for $\pi^0$-decay has been suggested to orginate in IC
scattering from a hard interstellar electron spectrum (e.g.,
\cite{PohlEsposito98}).

Fig.~\ref{fig9} (left) shows the \gray spectrum of the inner Galaxy for
a `conventional' case which matches the directly measured electron and
nucleon spectra and is consistent with synchrotron spectral index data
(\cite{MS98c}, \cite{SMR98}). Fig.~\ref{fig10} shows electron spectra
at $R_\odot = 8.5$ kpc in the disk for this model.  It fits the
observed \gray spectrum only in the range 30 MeV -- 1 GeV.
Fig.~\ref{fig9} (right) shows the case of $\pi^0$-decay \grays from a
hard nucleon spectrum (but still the `conventional' electron
spectrum).  This can improve the fit above 1 GeV but the high energy
antiproton and positron data probably exclude the hypothesis that the
local nucleon spectrum differs significantly from the Galactic average
(see Section \ref{positrons_pbar}).

\begin{figure}[t!]
   \centerline{
      \psfig{file=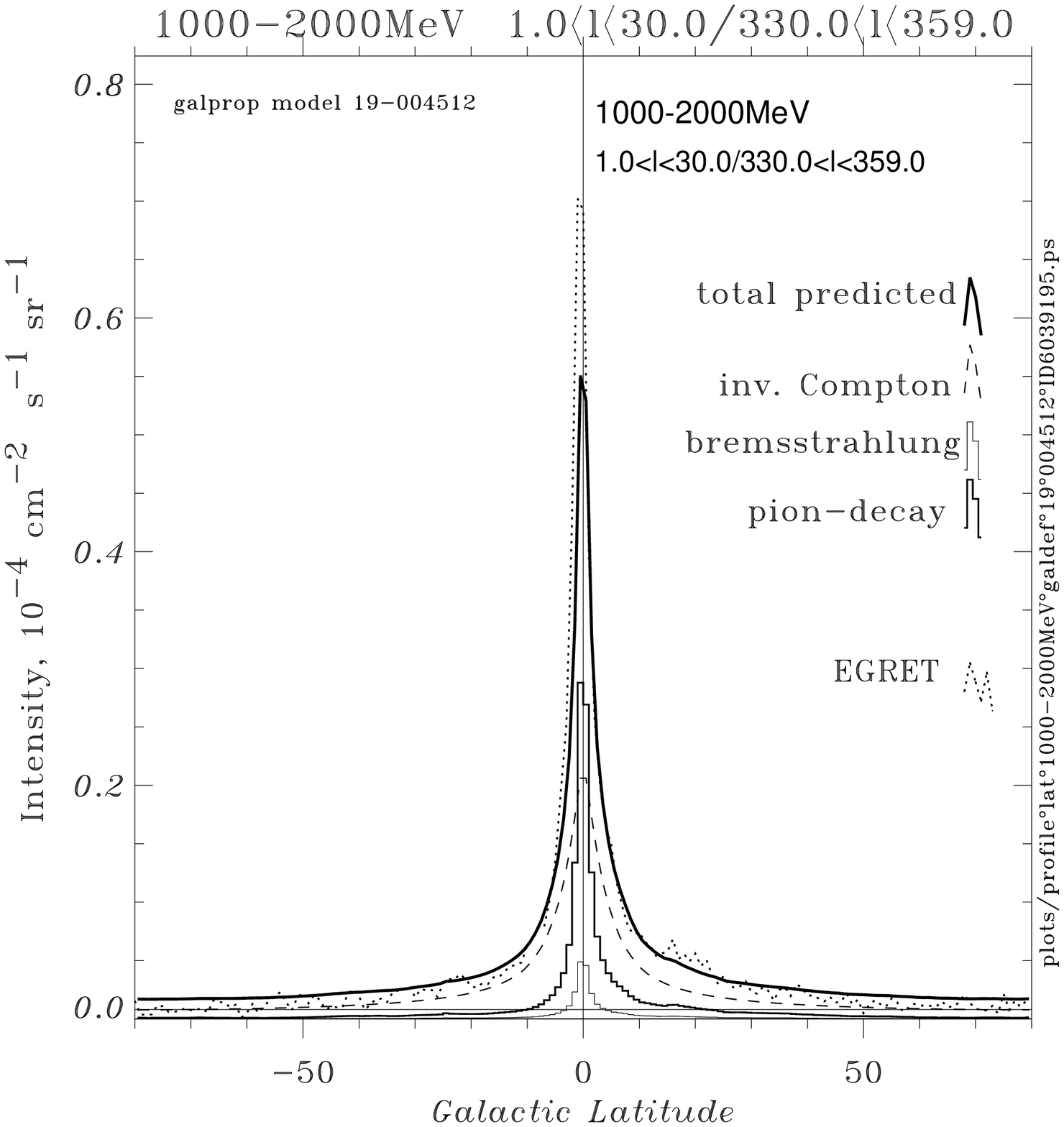,width=\fwb,clip=}
      \psfig{file=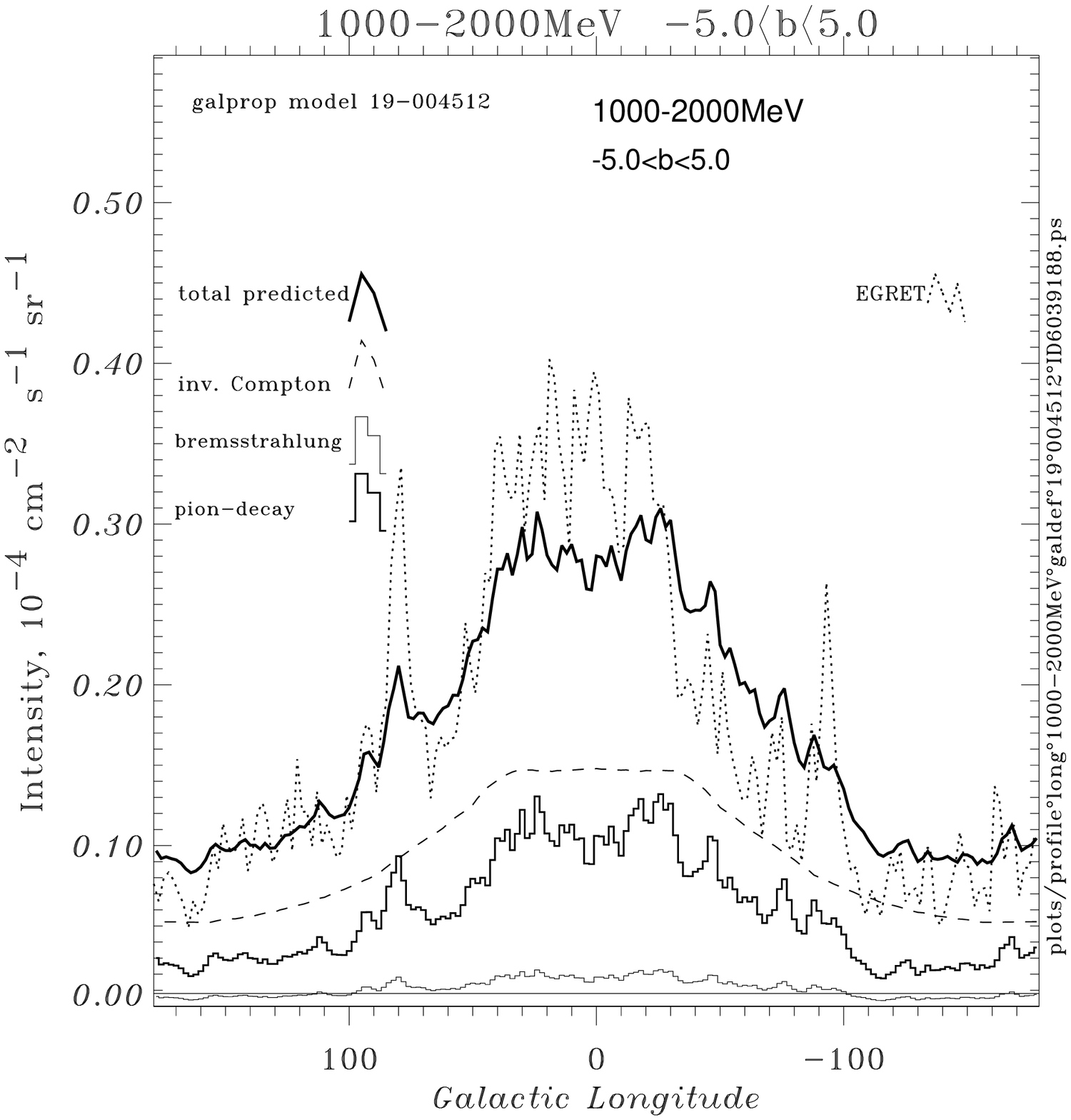,width=\fwb,clip=}
   }
\figcaption[fig11a.ps,fig11b.ps]{
Distributions of 1--2 GeV \grays computed for a hard electron spectrum
(reacceleration model) as compared to EGRET data (Cycles 1--4,
point sources removed, see \cite{SMR98}).  Contribution of various
components is shown as calculated in our model.  {\it Left panel:}
Latitude distribution ($330^\circ <l <30^\circ$).  {\it Right panel:}
Longitude distribution for $|b|<5^\circ$.
\label{fig11}}
\end{figure}

\placefigure{fig11}

We thus consider the `hard electron spectrum' alternative.  The
electron injection spectral index is taken as --1.7 (with
reacceleration), which after propagation provides consistency with
radio synchrotron data (a crucial constraint).   Following Pohl \&
Esposito (1998), for this model we do {\it not} require consistency
with the locally measured electron spectrum above 10 GeV since the
rapid energy losses cause a clumpy distribution so that this is not
necessarily representative of the interstellar average.  For this case,
the interstellar electron spectrum deviates strongly from that locally
measured as illustrated in Fig.~\ref{fig10}.  Because of the increased
IC contribution at high energies, the predicted \gray spectrum can
reproduce the overall intensity from 30 MeV -- 10 GeV but the detailed
shape above 1 GeV is still problematic.  Further refinement of this
scenario is presented in \cite{SMR98}.

Fig.~\ref{fig11} shows the model latitude and longitude \gray
distributions for the inner Galaxy for 1--2 GeV, convolved with the
EGRET point-spread function, compared to EGRET Phase 1--4 data (with
known point sources subtracted).  It shows that a model with large IC
component can indeed reproduce the data.  The latitude distribution
here is not as wide as at low energies owing to the rapid energy losses
of the electrons, so that an observational distinction between a
gas-related $\pi^0$-component from a hard nucleon spectrum and the IC
model does not seem possible on the basis of $\gamma$-rays alone, but
requires also other tests such as consistency with antiproton and
positron data (see Section~\ref{positrons_pbar}).

\begin{figure}[t!]
   \centerline{
      \psfig{file=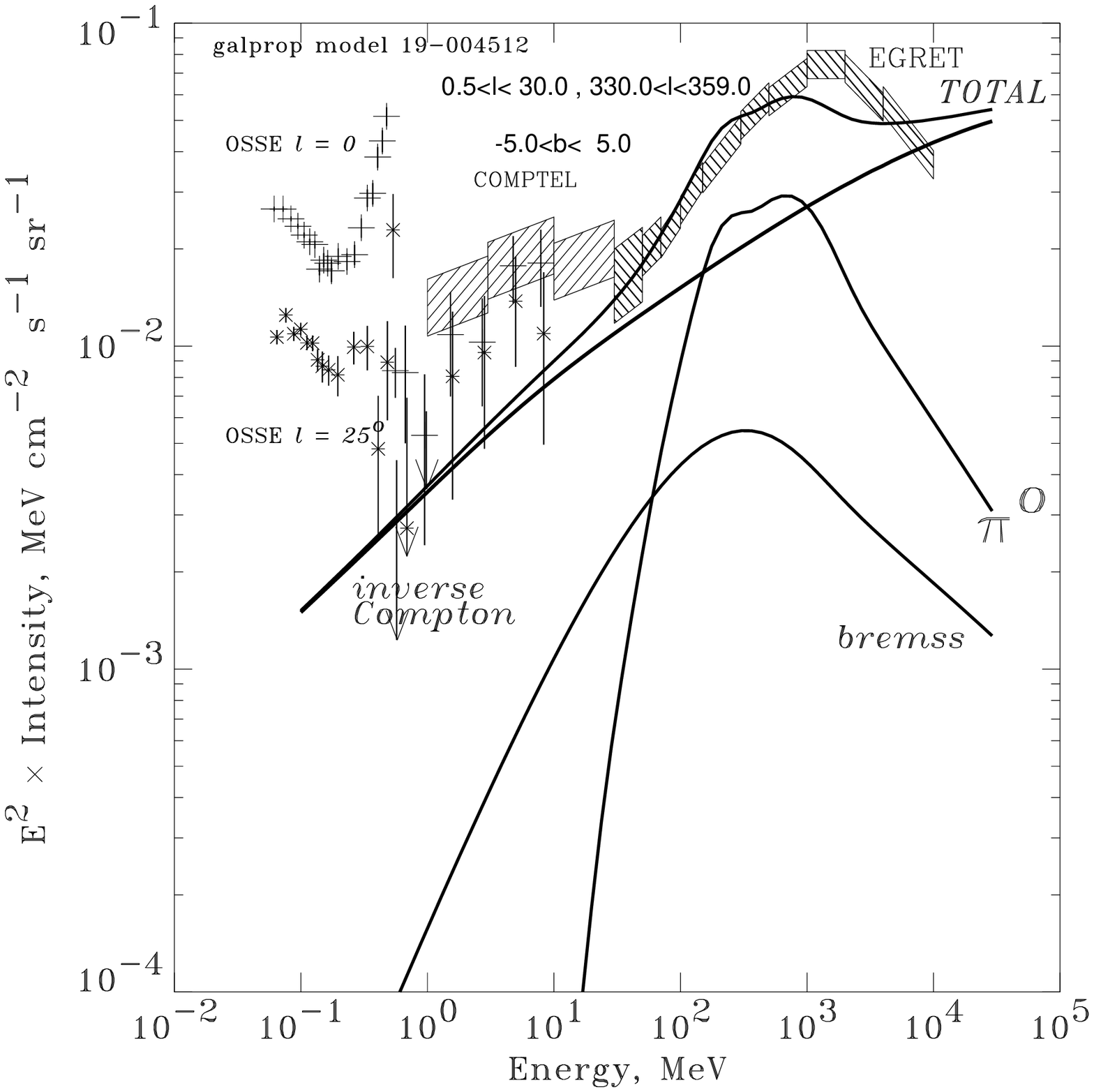,width=\fwb,clip=}
      \psfig{file=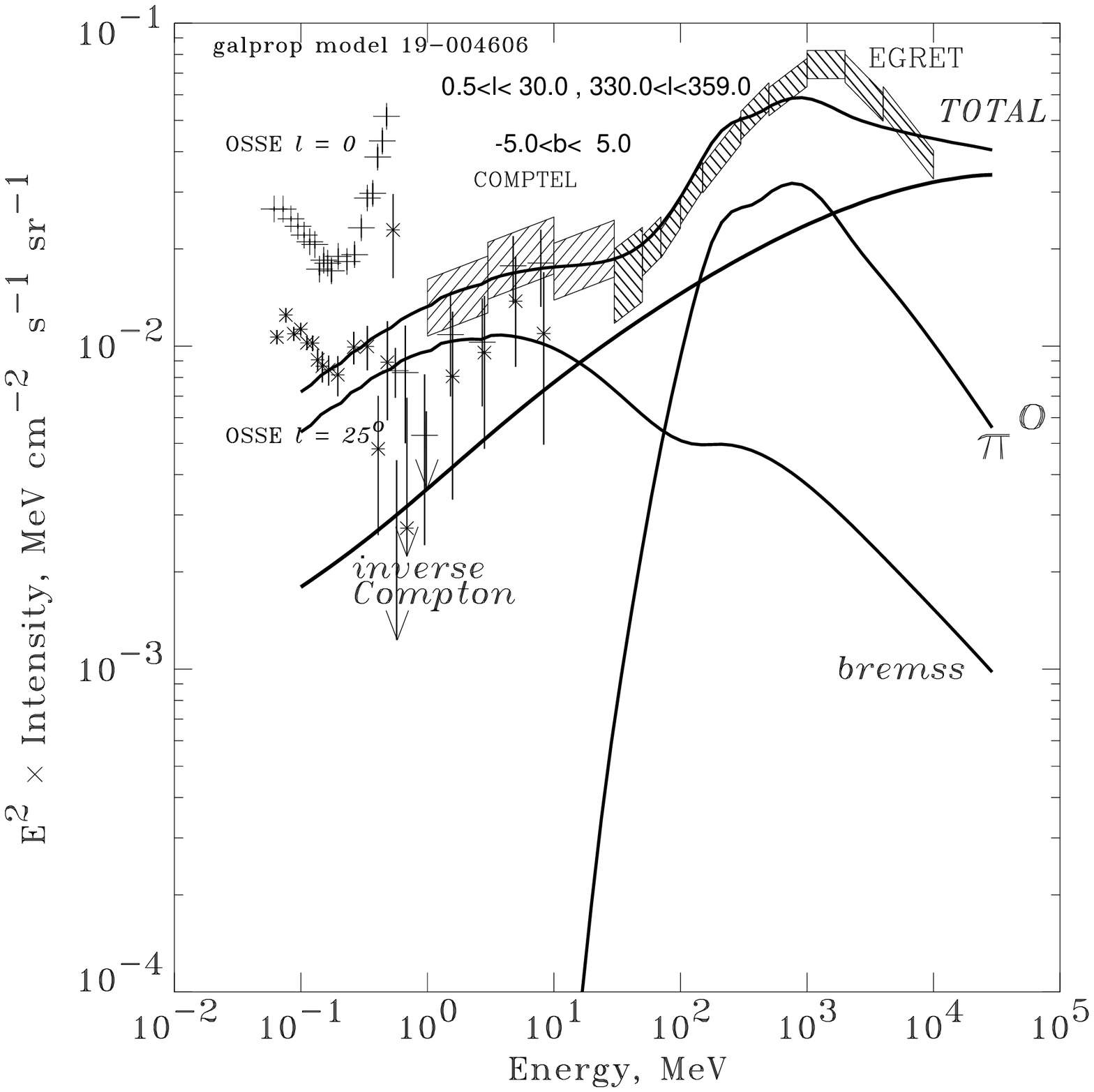,width=\fwb,clip=}
   }
   \caption[fig12a.ps,fig12b.ps]{
\gray spectrum of inner Galaxy compared to models with a hard electron
spectrum without (left) and with low-energy upturn (right).  Data as in
Fig.~\ref{fig9}.
\label{fig12}}
\end{figure}

\placefigure{fig12}

None of these models fits the \gray spectrum below $\sim$30 MeV as
measured by the Compton Gamma-Ray Observatory (Fig.~\ref{fig12} left).
In order to fit the low-energy part as diffuse emission
(Fig.~\ref{fig12} right), without violating synchrotron constraints
(\cite{SMR98}), requires a rapid upturn in the cosmic-ray electron
spectrum below 200 MeV (e.g., as in Fig.~\ref{fig10}).  However, in
view of the energetics problems (\cite{Skiboetal97}), a population of
unresolved sources seems more probable and would be the natural
extension of the low energy plane emission seen by OSSE
(\cite{Kinzer97}) and GINGA (\cite{Yamasaki97}).

\section{Conclusions}
Our propagation model has been used to study several areas of high
energy astrophysics.
We believe that combining information from classical cosmic-ray studies
with \gray and other data leads to tighter constraints on cosmic-ray
origin and propagation.

We have shown that simple diffusion/convection models have difficulty
in accounting for the observed form of the \BC\ ratio without special
assumptions chosen to fit the data, and do not obviate the need for an
{\it ad hoc} form for the diffusion coefficient.  On the other hand we
confirm the conclusion of other authors that models with reacceleration
account naturally for the energy dependence over the whole observed
range.  Combining these results points rather strongly in favour of the
reacceleration picture.

We take advantage of the recent Ulysses Be measurements to obtain
estimates of the halo size. Our limits on the halo height are
$4{\rm\ kpc} < z_h < 12$ kpc.  Our new limits should be an improvement
on previous estimates because of the more accurate Be data, our
treatment of energy losses, and the inclusion of more realistic
astrophysical details (such as, e.g., the gas distribution) in our
model.  The gradient of protons derived from \grays is smaller than
expected for SNR sources, and we therefore adopt a flatter source
distribution in order to meet the \gray constraints. This may just
reflect the uncertainty in the SNR distribution.

The positron and antiproton fluxes calculated are consistent with the
most recent measurements.  The $\bar{p}/p$ data point above 3 GeV and
positron flux measurements seem to rule out the hypothesis that the
local cosmic-ray nucleon spectrum differs significantly from the Galactic
average (by implication adding support to the `hard electron'
alternative), but confirmation of this conclusion must await more
accurate antiproton data at high energies.

Gamma-ray data suggest that the interstellar electron spectrum is
harder then that locally measured, but this remains to be confirmed by
detailed study of the angular distribution.
The low-energy Galactic \gray emission is difficult to explain as
truly diffuse and a point source population seems more probable.



\end{document}